\pdfoutput=1

\documentclass[twocolumn,showpacs,amssymb,aps,nofootinbib,floatfix,superscriptaddress]{revtex4-1}

\usepackage{epsfig}
\usepackage{hyperref}
\usepackage{comment}

\usepackage{graphicx,color}

\begin{document} 
\hbadness=10000

\title{Two-particle correlations in pseudorapidity in a hydrodynamic model}

\author{Piotr Bo\.zek}
\email{Piotr.Bozek@fis.agh.edu.pl}
\affiliation{AGH University of Science and Technology, Faculty of Physics and
Applied Computer Science, al. Mickiewicza 30, 30-059 Krakow, Poland}

\author{Wojciech Broniowski}
\email{Wojciech.Broniowski@ifj.edu.pl}
\affiliation{The H. Niewodnicza\'nski Institute of Nuclear Physics, Polish Academy of Sciences, 31-342 Krakow, Poland}
\affiliation{Institute of Physics, Jan Kochanowski University, 25-406 Kielce, Poland}

\author{Adam Olszewski}
\email{Adam.Olszewski.fiz@gmail.com}
\affiliation{Institute of Physics, Jan Kochanowski University, 25-406 Kielce, Poland}

\begin{abstract}
Two-particle pseudorapidity correlations of  hadrons produced in  Pb+Pb collisions at \mbox{$\sqrt{s_{NN}}=2.76$~TeV} at the CERN Large Hadron Collider are analyzed in the framework of
a model based on viscous 3+1-dimensional hydrodynamics with the Glauber initial condition.   
Based on our results, we argue that the correlation from resonance decays, formed at a late stage of the evolution, produce significant effects. In particular, their contribution
to the event averages of the coefficients of the expansion in the Legendre basis explain 60-70\% of the experimental values. 
We have  proposed an accurate way to compute these  
coefficients, independent of the binning in pseudorapidity, 
and tested a double expansion of the two-particle correlation function in the azimuth and pseudorapidity, which allows us
to investigate the pseudorapidity correlations between harmonics of the collective flow. In our model, these quantities are also dominated by 
non-flow effects from the resonance decays.
Finally, our method can be used to compute higher-order cumulants for the expansion in orthonormal polynomials \cite{Bzdak:2015dja} which 
offers a suitable way of eliminating the non-flow effects from the correlation analyses.
\end{abstract}

\date{14 September 2015}

\pacs{25.75.-q, 25.75Gz, 25.75.Ld}

\keywords{relativistic heavy-ion collisions, pseudorapidity correlations,  harmonic flow, torque effect}

\maketitle


\section{Introduction \label{sec:intro}}

The mechanism of energy deposition in relativistic nuclear collisions is a subject of intense studies. 
Whereas most of the investigations are concerned with the entropy-deposition profile in the transverse plane and the 
resulting transverse expansion (for reviews see, e.g.,~\cite{Heinz:2013th,Gale:2013da}), the dynamics in the longitudinal direction is less explored, and
has been recently gaining more attention with the new experimental analyses from the CERN Large Hadron Collider (LHC) expected shortly.
Such studies could give valuable insight into the initial energy and momentum distributions in rapidity~\cite{Dusling:2009ni,*Armesto:2006bv,*Fukushima:2008ya},  the longitudinal
collective dynamics~\cite{Bozek:2007qt,*Ryblewski:2012rr,*Martinez:2010sc,*Casalderrey-Solana:2013sxa}, or hydrodynamic fluctuations~\cite{Kapusta:2011gt,*Gavin:2011gr}. 
Correlations in (pseudo)rapidity can be studied in various ways, in particular, as correlations of the transverse 
flow at different rapidity bins, or as multiplicity correlations in rapidity. The first case requires an intermediate collective expansion stage producing 
flow~\cite{Bozek:2010vz,Petersen:2011fp,*Xiao:2012uw,*Jia:2014ysa,*Jia:2014vja,*Pang:2014pxa,*Khachatryan:2015oea,Bozek:2015bna,Bozek:2015bha}, whereas the 
particle distribution and multiplicity correlations  in rapidity
are not modified significantly during the fireball expansion, thus are expected to reflect more closely the initial conditions in the fireball.
In other words, the multiplicity correlations arise even without any collective expansion. 

Correlations of the multiplicity of particles observed in high energy collisions in different pseudorapidity intervals
have been studied  in a number of colliding systems~%
\cite{Bzdak:2012tp,Dusling:2009ni,Back:2006id,*Bzdak:2009xq,*Abelev:2009ag,*Feofilov:2013kna,*De:2013bta,%
*Amelin:1994mf,*Braun:1997ch,*Brogueira:2006yk,*Yan:2010et,*Bialas:2011xk,*Olszewski:2013qwa,*Olszewski:2015xba}.
The most common approach is based on 
the correlation of the number of particles in forward and a backward pseudorapidity bins, $\langle n_F n_B\rangle$,
or related observables. 

Bzdak and Teaney have proposed to expand the two-point correlation function in pseudorapidity in
a basis of orthogonal polynomials~\cite{Bzdak:2012tp}. The correlations
are then written in terms of the corresponding expansion coefficients $\langle a_n a_m \rangle $. The extracted coefficients 
can serve to parametrize  event by event fluctuations of the particle distribution in pseudorapidity.
A basis of the Legendre polynomials \cite{Jia:2015jga} has  been used for the expansion of the correlation in pseudorapidity
for the case of Pb+Pb collisions at $\sqrt{s}=2760$~GeV, recently measured by the \mbox{ATLAS} Collaboration~\cite{ATLAS:2015kla,*AnmTALK}.

In this work we present predictions of the relativistic hydrodynamic model for the two-particle correlations 
in pseudorapidity, focusing on correlations generated in the late stage of the collision via resonance decays. 
Our approach consists of a Glauber Monte Carlo model with asymmetric longitudinal emission profile for the initial state, and the viscous 
3+1D hydrodynamic evolution of the  fireball, followed by statistical hadron emission at freeze-out.
Our main result is that the late-stage correlations from resonance decays contribute largely (about a half of the measured 
values) to the correlations extracted in terms of the $\langle a_n a_m \rangle$ coefficients. The missing strength 
should be attributed to the correlations generated in the earlier stages of the evolution (initial state, jets).

\section{Two-particle correlation \label{sec:corr}}

The two-particle correlation in pseudorapidity, scaled by the one-particle distributions,  is defined as
\begin{equation}
C(\eta_1,\eta_2)= \frac{\langle N(\eta_1)N(\eta_2) \rangle -\langle N(\eta_1)\rangle \delta(\eta_1-\eta_2)}{\langle N(\eta_1)\rangle \langle N(\eta_2)\rangle} \label{eq:C2} \ ,
\end{equation}
where $N(\eta)$ denotes the distribution of the number of hadrons at $\eta$ and the averaging is over events in a selected centrality class.
In the experiment, the correlation function is constructed as the ratio of the histogram for particle pairs from 
physical events to the histogram constructed from mixed events in the same centrality class~\cite{ATLAS:2015kla}
\begin{equation}
C(\eta_1,\eta_2)=\frac{S(\eta_1,\eta_2)}{B(\eta_1,\eta_2)} \ .
\label{eq:c2exp}
\end{equation}
 
We note that the definition~ (\ref{eq:C2}) corresponds to the scaled second factorial moment of the multiplicity distribution,
which depends on the centrality definition and the width of the centrality bin. 
To reduce the effects of the overall multiplicity fluctuations, the  ATLAS collaboration uses a modified correlation function
\begin{equation}
C_N(\eta_1,\eta_2)=\frac{C(\eta_1,\eta_2)}{C_p(\eta_1)C_p(\eta_2)} \ , 
\label{eq:CN}
\end{equation}
with $C_p(\eta_1)=\frac{1}{2Y}\int_{-Y}^Y C(\eta_1,\eta_2) d \eta_2$.
The experimental analysis suggests that  $C_N(\eta_1,\eta_2)$ 
is approximately independent of the definition of centrality~\cite{Jia:2015jga,ATLAS:2015kla}.

\section{Expansion in orthonormal polynomials \label{sec:poly}}

As the shape of the distribution function ${N(\eta)}/{\langle N(\eta) \rangle}$ fluctuates event by event, it can be expanded
in a basis of orthogonal functions~\cite{Bzdak:2012tp}
\begin{equation}
\frac{N(\eta)}{\langle N(\eta) \rangle}= 1 + \sum_{n=0}^\infty a_n T_n\left(\frac{\eta}{Y}\right) \ .
\label{eq:aexp}
\end{equation}
  For the case of  Legendre polynomials $P_n(x)$, the normalized functions are
$T_n(\frac{\eta}{Y})=\sqrt{\frac{2n+1}{2}}P_n(x)$ \cite{Jia:2015jga},
where $[-Y,Y]$ is the  pseudorapidity range on which the correlation functions are measured, such that the orthonormality 
condition takes the form
\begin{equation}
 \int_{-Y}^Y T_n\left( \frac{\eta}{Y} \right)  T_m\left( \frac{\eta}{Y} \right) \frac{d\eta}{Y} = \delta_{nm} \ .
\end{equation}

The event-average $\langle a_n a_m \rangle$ can be calculated from the two-particle correlation function
\begin{equation}
\langle a_n a_m \rangle = \int_{-Y}^Y \frac{d \eta_1}{Y} \int_{-Y}^Y \frac{d \eta_2}{Y}  C(\eta_1,\eta_2) T_n\left(\frac{\eta_1}{Y}\right) 
T_m\left(\frac{\eta_1}{Y}\right) \label{eq:aaint} \ .
\end{equation}
The procedure is rather complicated, as first the two-particle correlation function must be constructed with sufficiently fine binning. In the case of 
low statistics, large binning of $C$ introduces biases. 

The estimate of the integral (\ref{eq:aaint}) can be simply obtained from
\begin{eqnarray}
\langle a_n a_m \rangle & = & \left\langle \sum_{a\neq b}  \frac{T_n\left(\frac{\eta_a}{Y}\right)}{Y \langle N(\eta_a) \rangle}
 \frac{T_m\left(\frac{\eta_b}{Y}\right)}{Y \langle N(\eta_b) \rangle}  \right\rangle  \label{eq:anm2} \\
 & =& \left\langle   {\sum_{a} \frac{T_n\left(\frac{\eta_a}{Y}\right)}{Y \langle N(\eta_a) \rangle} \sum_{b} \frac{T_m\left(\frac{\eta_b}{Y}\right)}{Y \langle N(\eta_b) \rangle}} \right \rangle \nonumber \\ & -&
 \left \langle \sum_{a} \frac{T_n\left(\frac{\eta_a}{Y}\right)}{Y \langle N(\eta_a) \rangle} \frac{T_m\left(\frac{\eta_a}{Y}\right)}{Y \langle N(\eta_a) \rangle}  \right\rangle \  , \nonumber
\end{eqnarray}
where the sums are over hadrons in the given event and the averages are over events. Equation (\ref{eq:anm2}) produces very stable results, free of the binning bias.

In the experimental analysis of Ref.~\cite{ATLAS:2015kla} the function $C_N(\eta_1,\eta_2)$ instead of  $C(\eta_1,\eta_2)$ is used in Eq.~(\ref{eq:aaint}). We have checked that 
in our case the resulting difference for the $\langle a_n a_m \rangle$ coefficients for $1 \le n, m \le 5$ is very small, a fraction of percent,\footnote{A correction, which is tiny, could be worked out 
along the lines of Ref.~\cite{Jia:2015jga}.} hence in the following we will use 
$C(\eta_1,\eta_2)$ in Eq.~(\ref{eq:anm2}). In addition, the function $C(\eta_1,\eta_2)$ is, in the experiment, normalized to 1.
To conform to this convention we rescale the coefficients obtained from Eq.~(\ref{eq:anm2}):
\begin{equation}
\langle a_n a_m \rangle \to \frac{\langle a_n a_m \rangle}{1+\langle a_0 a_0 \rangle/2} \ .
\end{equation} 
In practice, for centrality bins in the model 
calculation defined by the number of participant nucleons, the correction to the normalization is less than 2\%.

The motivation of the studies of Ref.~\cite{Bzdak:2012tp,Jia:2015jga,ATLAS:2015kla} was to transform the two-particle distributions into 
a series of coefficients $\langle a_n a_m \rangle$ with a simple interpretation.
For instance,  the coefficient $\langle a_1 a_1\rangle$ is related to the asymmetry in the 
entropy deposition in rapidity from the forward and backward going participant nucleons. 
The asymmetry of the deposition in rapidity is visible in the charged particle distribution 
in pseudorapidity in asymmetric collisions~\cite{Bialas:2004su} and in forward-backward multiplicity distributions~\cite{Bzdak:2009xq}.

\section{Results from the hydrodynamic model \label{sec:hydro}}

\begin{figure}[tb]
\includegraphics[width=0.45 \textwidth]{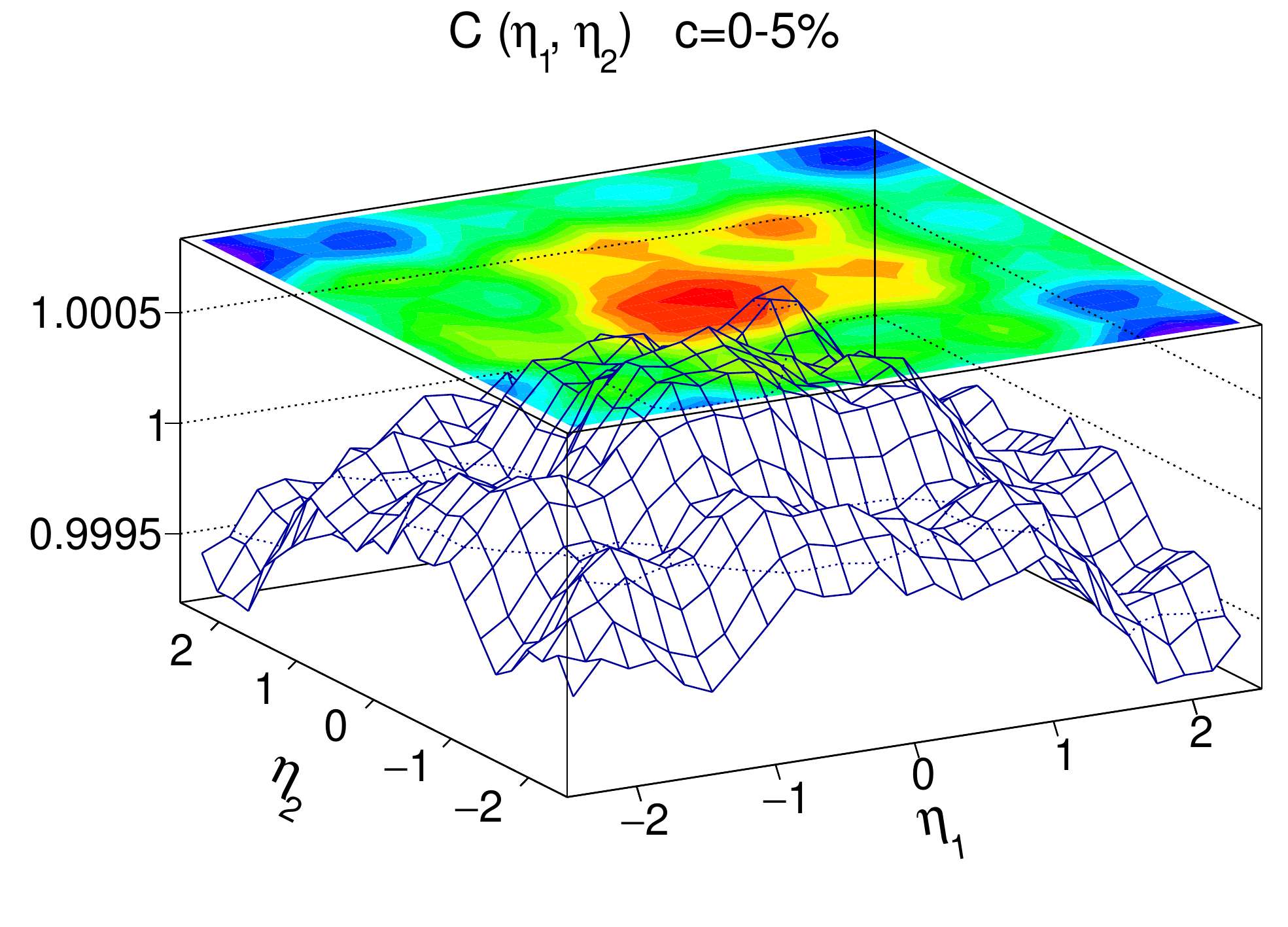}\\ 
\includegraphics[width=0.45 \textwidth]{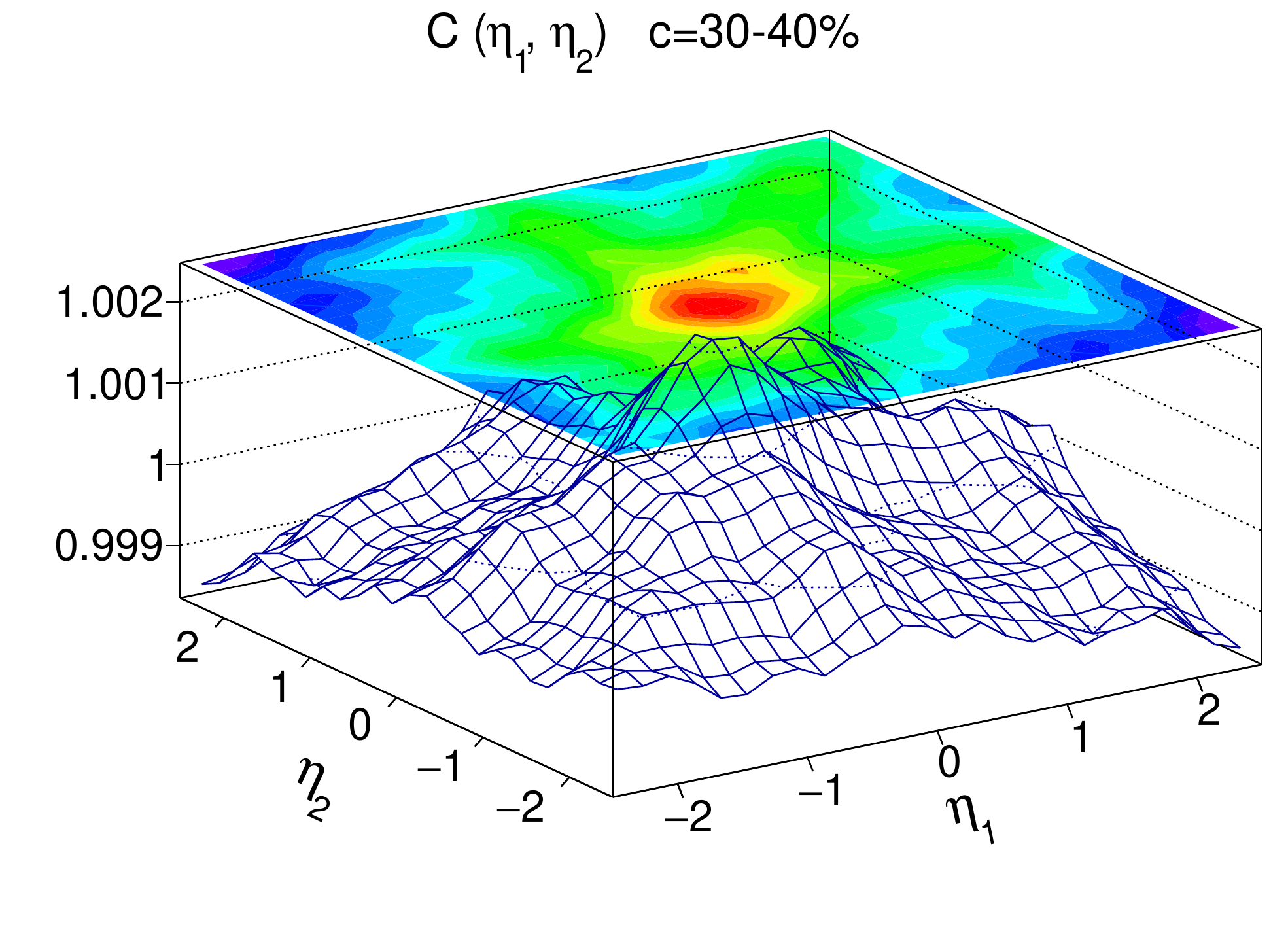}\\ 
\includegraphics[width=0.45 \textwidth]{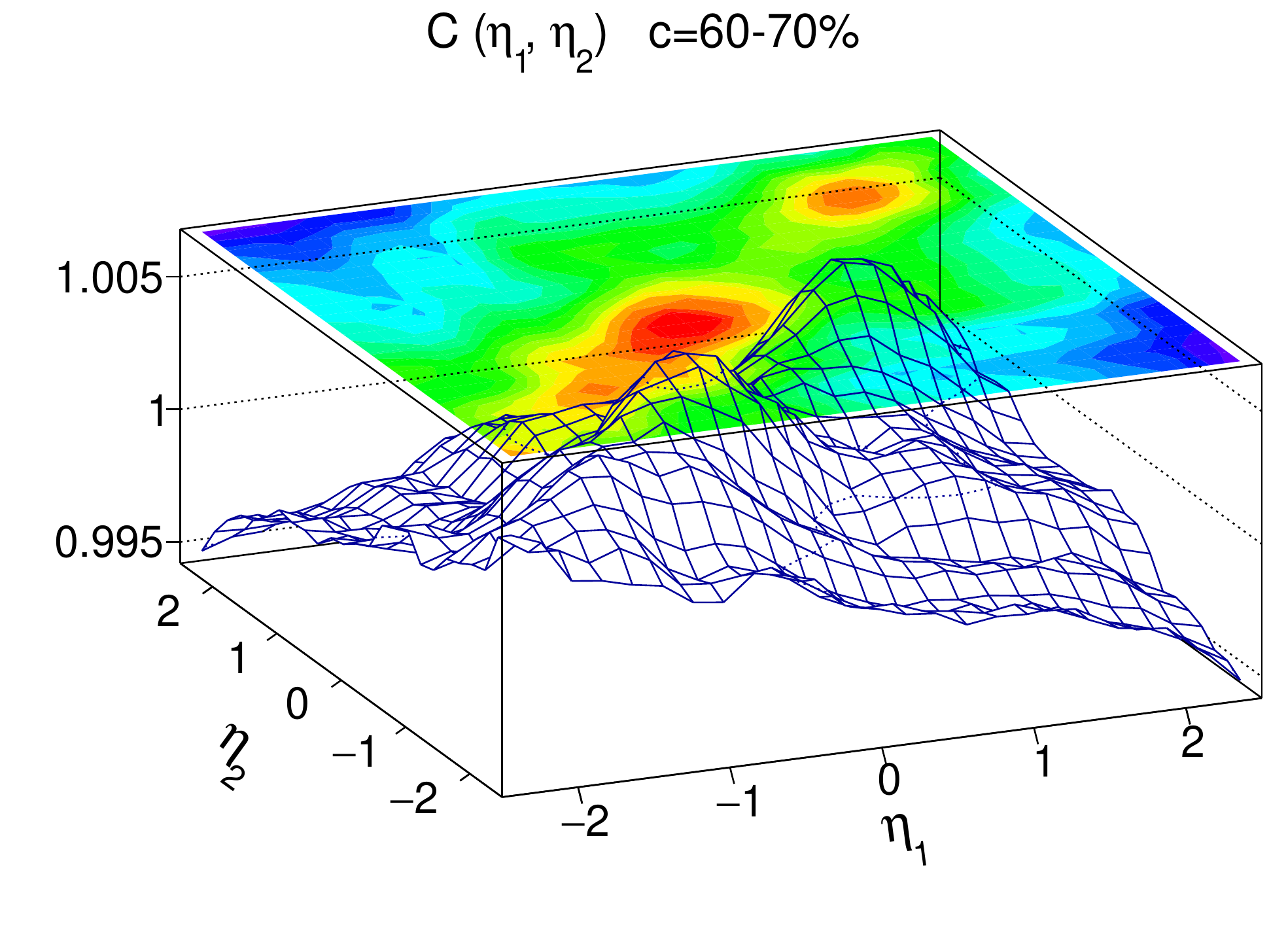} 
\caption{(color online) Two dimensional correlation function in pseudorapidity for charged particles in Pb+Pb collisions at \mbox{$\sqrt{s}=2.76$~TeV}
at three centrality classes 0-5\%, 30-40\% and 60-70\% in panels (a), (b) and (c), respectively.
\label{fig:cor}}
\end{figure}

We use the 3+1-dimensional viscous hydrodynamics~\cite{Bozek:2011ua} to model the evolution of the fireball created in Pb+Pb collisions
at $\sqrt{s}=2.76$~TeV. The initial entropy density in the transverse plane  is calculated in GLISSANDO~\cite{Broniowski:2007nz,*Rybczynski:2013yba}, implementing the Glauber Monte Carlo model.
The initial profile in the longitudinal direction (in space-time rapidity) is considered in two qualitatively different scenarios.
In the first scenario, the entropy distribution in space-time rapidity from a left- and right-going participant nucleons is
of the form
\begin{equation}
f_{\pm}(\eta_\parallel)= \frac{\eta_{\rm beam}\pm \eta_\parallel}{ y_{\rm beam}} H(\eta_\parallel)\  \mbox {\rm for } \ |\eta_\parallel|<y_{\rm beam} \  ,
\label{eq:lprof}
\end{equation}
where
\begin{equation}
H(\eta_\parallel)=\exp\left(-\frac{(|\eta_\parallel|-\eta_p)^2\Theta(|\eta_\parallel|-\eta_p)}{2\sigma_\eta^2}\right) ,
\end{equation}
and $y_{\rm beam}$ is the rapidity of the beam. For the LHC, the parameters determining the shape are $\sigma=1.4$ and $\eta_p=2.4$. 
The asymmetric distribution of the deposited entropy between the forward and backward rapidity hemisphere leads, together with the fluctuations in the number of participants, lead
to nontrivial correlations between forward and backward rapidity bins, both in multiplicity~\cite{Bzdak:2009xq}
and in the flow angle orientation~\cite{Bozek:2010vz,*Bozek:2015bha}.  The latter has been termed the {torque} effect, hence we label our calculations based on Eq.~(\ref{eq:lprof}) as {\em torque}.

The reference scenario assumes that the initial entropy profile in space-time rapidity is symmetric,
\begin{equation}
f_{\pm}(\eta_\parallel)=  H(\eta_\parallel)\  \mbox {for } \ |\eta_\parallel|<y_{\rm beam} \ .
\label{eq:lprof}
\end{equation}
In that case (labeled {\em no torque}) in each event the fireball density has a backward-forward symmetry, hence no shape fluctuations of odd reflection symmetry are possible. 
To summarize, the torque case includes certain initial-state fluctuations in rapidity, while the no-torque case does not.

At freeze-out, hadrons are emitted, and later resonance decays occur. 
The decays of resonances introduce short-range correlations of length of about one unit in pseudorapidity,  
leading to a nontrivial structure of the two-dimensional correlation functions. Note that
another source of correlation in the late stage, unrelated 
to the fireball shape fluctuations, is due to local charge conservation~\cite{Jeon:2001ue,Bozek:2012en}.

\begin{figure}[tb]
\includegraphics[width=0.45 \textwidth]{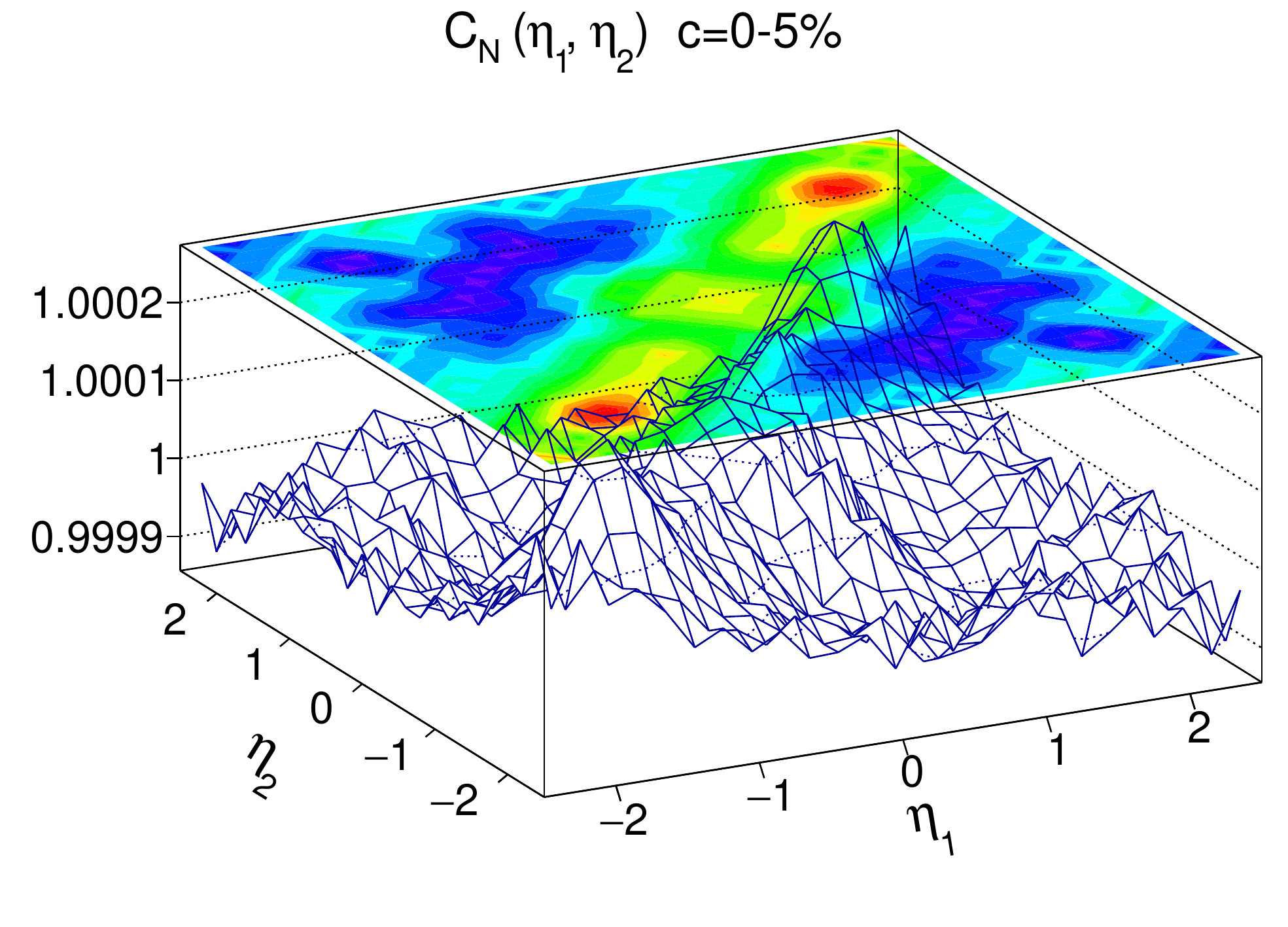}\\ 
\includegraphics[width=0.45 \textwidth]{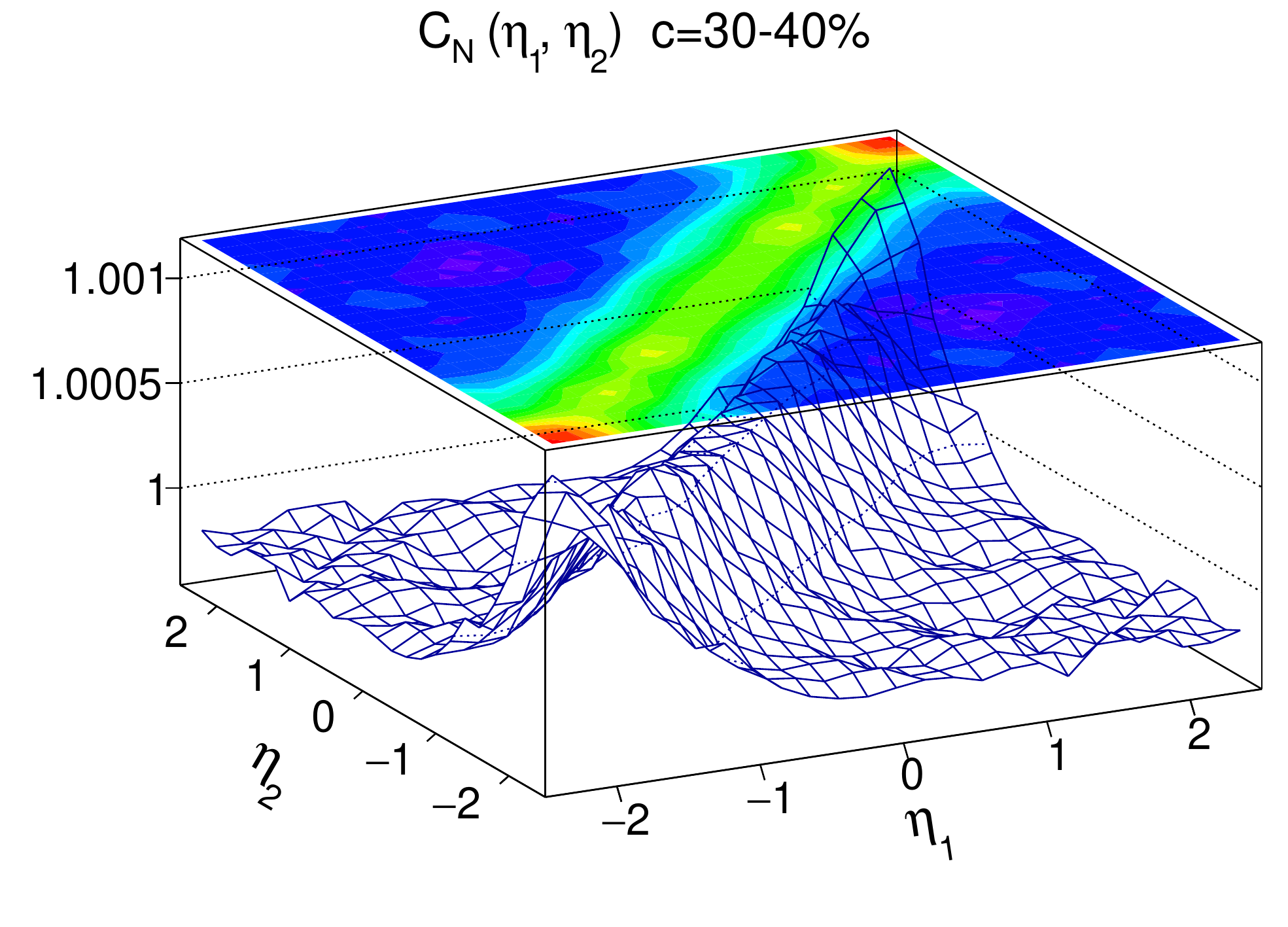}\\ 
\includegraphics[width=0.45 \textwidth]{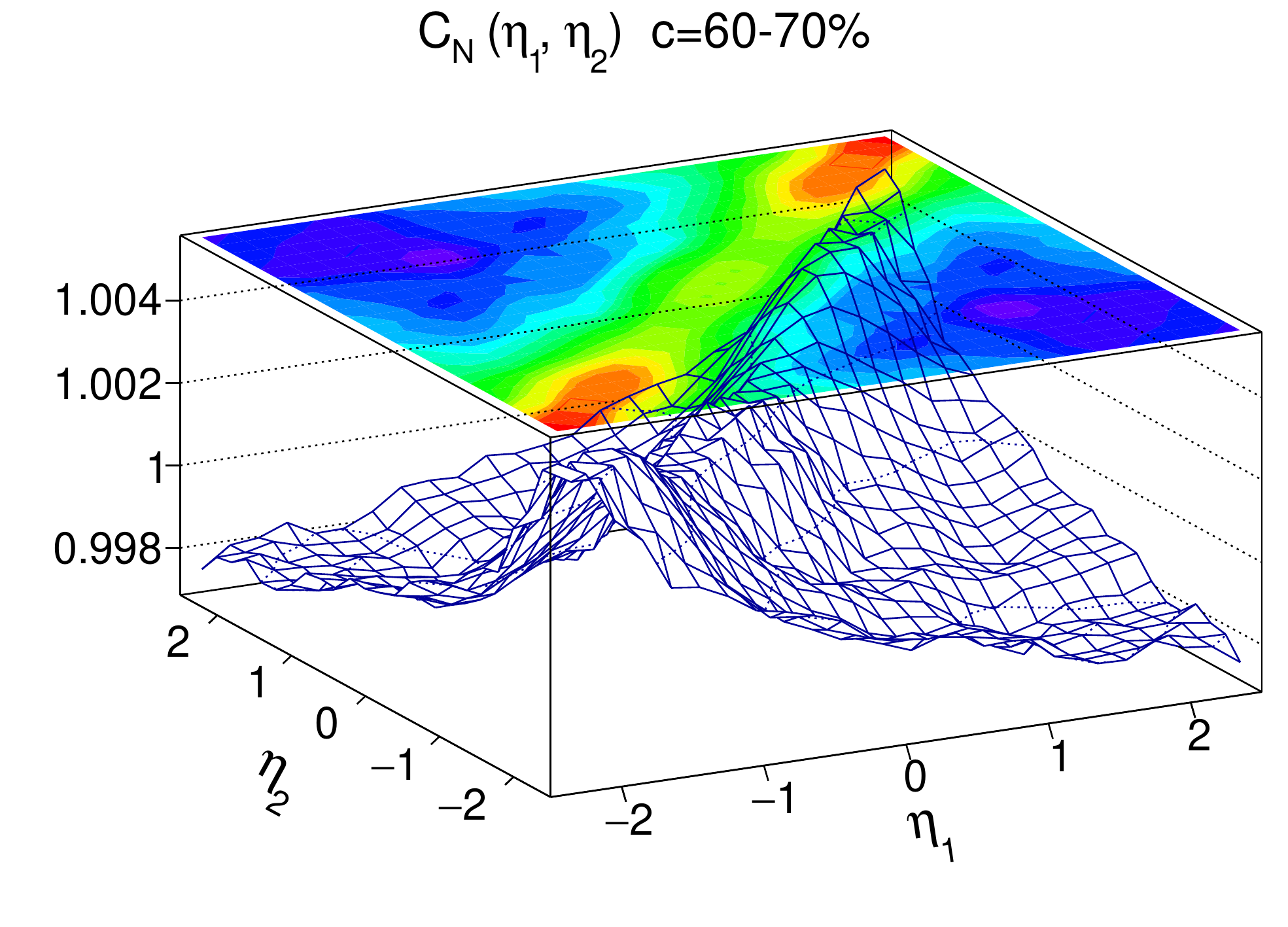} 
\caption{(color online)  Same as Fig. \ref{fig:cor} but for the corrected correlations function $C_N(\eta_1,\eta_2)$ (\ref{eq:CN}).
\label{fig:CN}}
\end{figure}

The correlation function (\ref{eq:c2exp}) is calculated from realistic, finite-multiplicity events, generated after the hydrodynamic evolution 
with THERMINATOR~\cite{Kisiel:2005hn,*Chojnacki:2011hb}. We use the freeze-out temperature $T_f=150$~MeV. 
The simulated events include the short-range correlations from resonance decays. In Fig.~\ref{fig:cor} we show the two-dimensional correlations 
for three different centrality classes. Charged particles with $p_\perp>0.5$~GeV and $|\eta|<2.5$ are taken to simulate the ATLAS acceptance. 

For $C(\eta_1,\eta_2)$, plotted in Fig.~\ref{fig:cor} for three sample centralities, a peak from the short range 
correlations is clearly visible around $\eta_1 \simeq\eta_2$. When passing to $C_N(\eta_1,\eta_2)$, we note that the 
denominator in Eq.~(\ref{eq:CN}) is smaller than one at large $|\eta_{1,2}|$, hence it causes relative enhancement of the correlation measure in this region. 
As a result, the shape of the correlation function is changed
significantly when passing from $C(\eta_1,\eta_2)$ to $C_N(\eta_1,\eta_2)$, cf.~Figs.~\ref{fig:cor} and \ref{fig:CN}. In particular, a saddle-like form appears, corresponding to a term of the form
$A (\eta_+^2-\eta_-^2)$, where $\eta_\pm=\eta_1 \pm  \eta_2$. Such a term is expected from event-by-event asymmetry of the 
initial distribution function~\cite{Bzdak:2012tp}, giving a nonzero value of $\langle a_1 a_1 \rangle$. Without this asymmetry, the short-range correlations 
are expected to be a function of $|\eta_1-\eta_2|$ only~\cite{Xu:2012ue,*Xu:2013sua}.

However, in our simulations almost the same value of  $\langle a_1 a_1 \rangle$
is obtained from the correlation functions $C(\eta_1,\eta_2)$ and $C_N(\eta_1,\eta_2)$.
It thus suggests that the observed dependence of the correlation function on $\eta_\pm$ does not directly prove the 
existence  of correlations induced by the event-by-event fluctuations of  the distribution. 
  
\begin{figure}[tb]
\includegraphics[width=0.45 \textwidth]{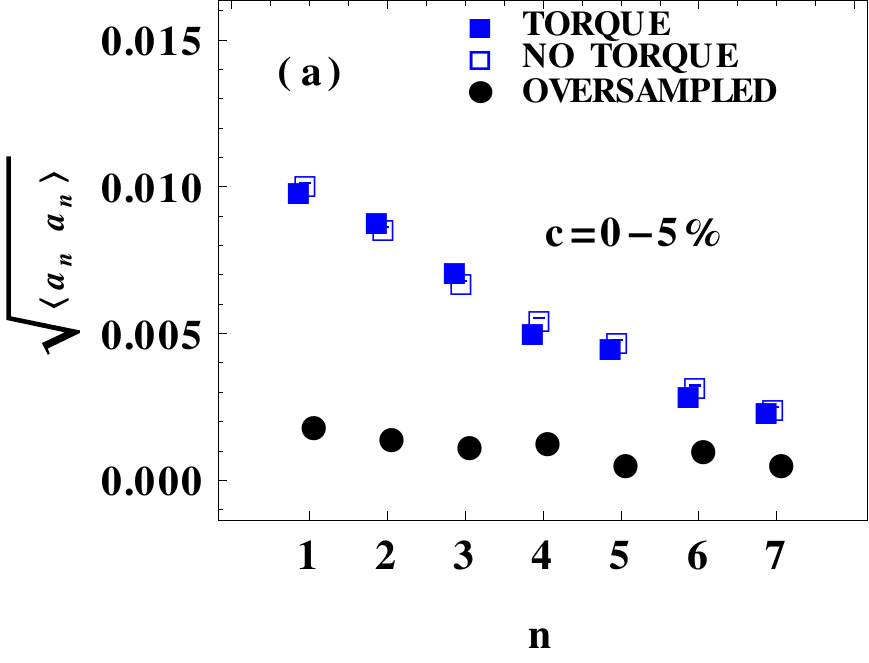}\\ 
\vspace{-8mm}
\includegraphics[width=0.45 \textwidth]{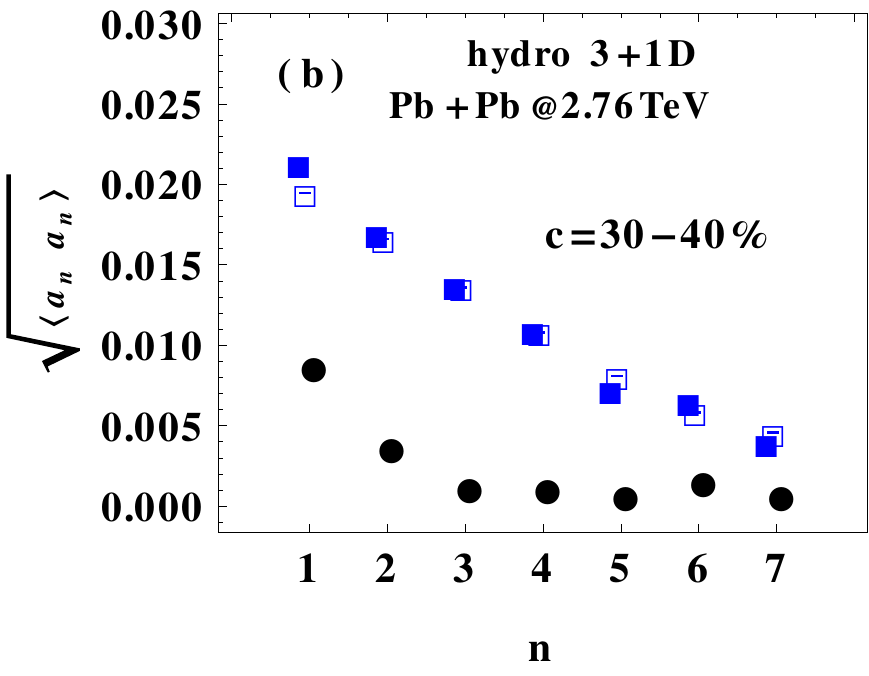}
\caption{(color online)  Calculated coefficients $\sqrt{\langle a_n a_n \rangle}$ for centrality $0-5$\% (panel a) and $30-40$\% (panel b)
for the torque and no-torque models, as well as for the oversampled events for the torque case (see text for details).}
\label{fig:ann}
\end{figure}

\begin{figure}[tb]
\includegraphics[width=0.45 \textwidth]{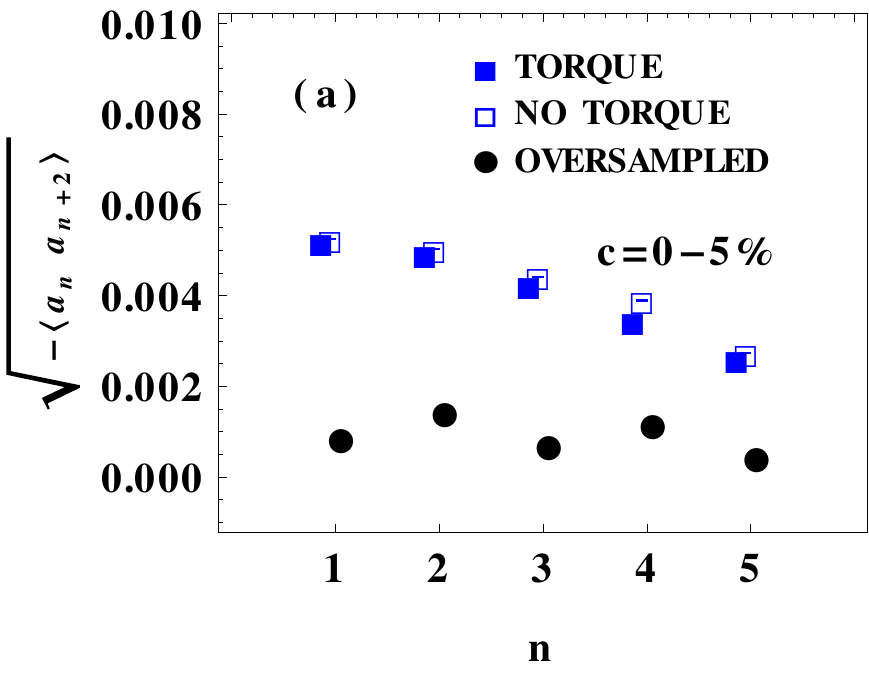}\\ 
\vspace{-8mm}
\includegraphics[width=0.45 \textwidth]{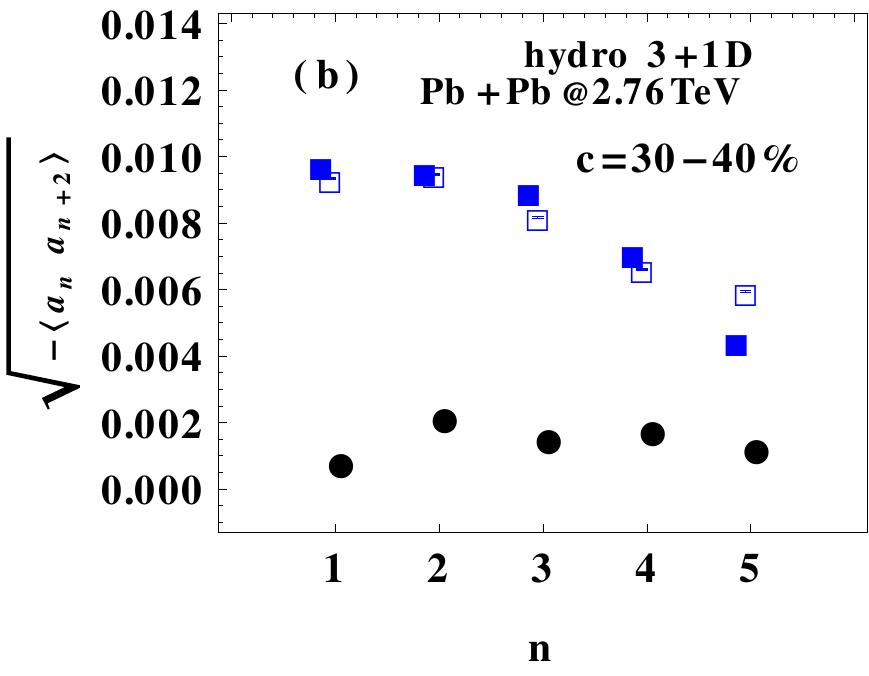}
\caption{(color online)  Same as in Fig. \ref{fig:ann} but for the the coefficients $\sqrt{ - \langle a_n a_{n+2}  \rangle}$.
\label{fig:ann2}}
\end{figure}

In Fig.~\ref{fig:ann} we show the calculated coefficients $\sqrt{\langle a_n a_n \rangle }$, $n=1,\dots,7$ for two sample centralities.
The magnitude predicted by the model reaches about 60-70\% of the values observed experimentally~\cite{ATLAS:2015kla}. The trend of the dependence 
on the rank $n$ is similar as in the experiment. Similar conclusions can be made for the non-diagonal coefficients 
$\sqrt{ -\langle a_n a_{n+2} \rangle}$ shown in Fig.~\ref{fig:ann2}. With the available statistics, we cannot  calculate higher order averages $\langle a_n a_{n+4} \rangle$.
The results for the two scenarios of the the initial conditions, torque and no-torque, are shown. Interestingly, both calculations give very similar results.
This shows that in our model the dominant contribution in the observed 
signal comes from the short-range correlations due to resonance decays.

\section{Double expansion of correlations functions in azimuthal angle and pseudorapidity}

\begin{figure}[tb]
\includegraphics[width=0.45 \textwidth]{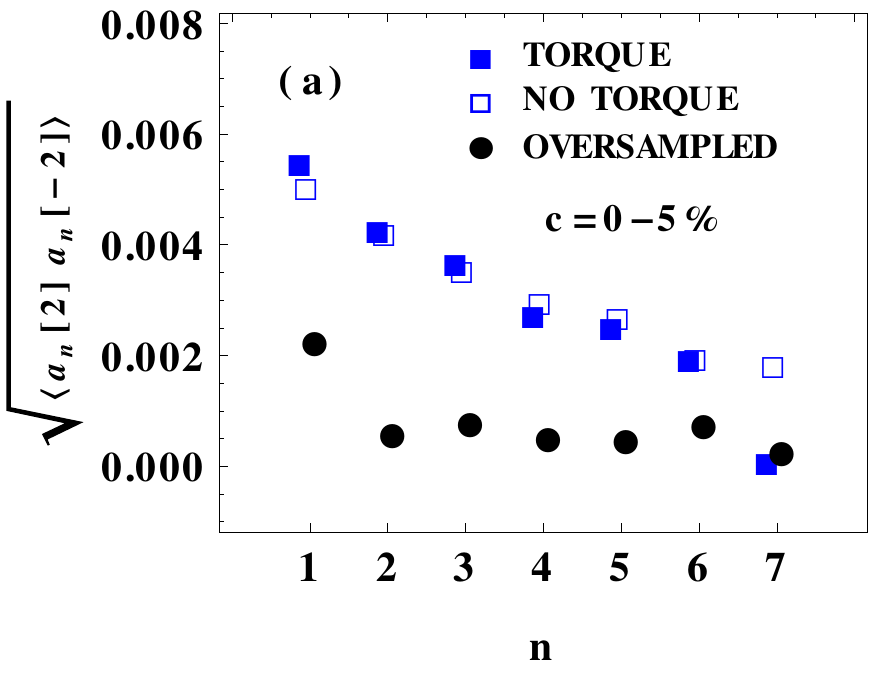}\\
\vspace{-8mm}
\includegraphics[width=0.45 \textwidth]{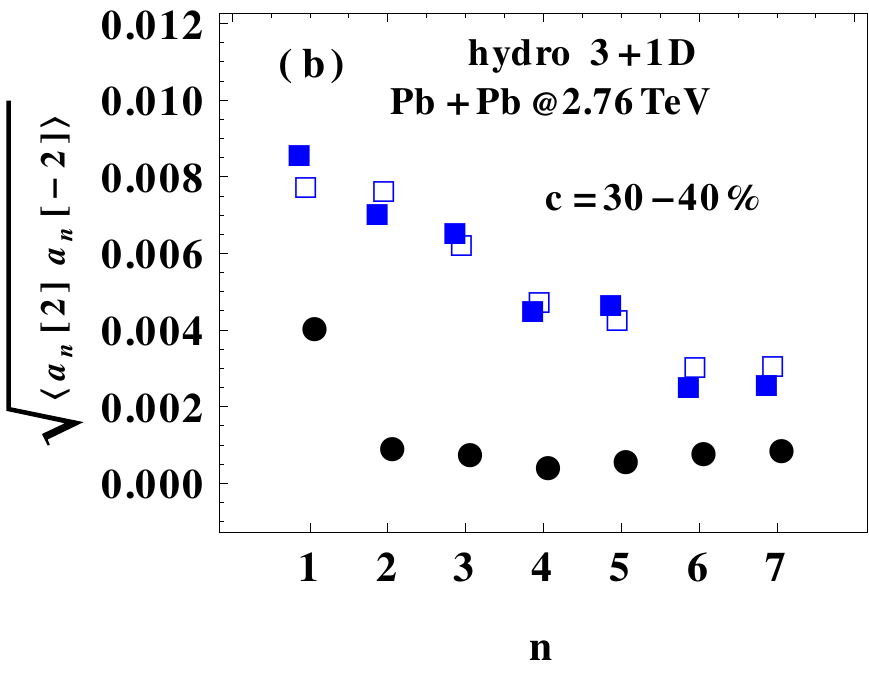} 
\caption{(color online)  
Same as in Fig. \ref{fig:ann} but for the the coefficients of the second order harmonic $\sqrt{\langle a_n[2] a_n[-2]\rangle}$.
\label{fig:annv2}}
\end{figure}
\begin{figure}[tb]
\includegraphics[width=0.45 \textwidth]{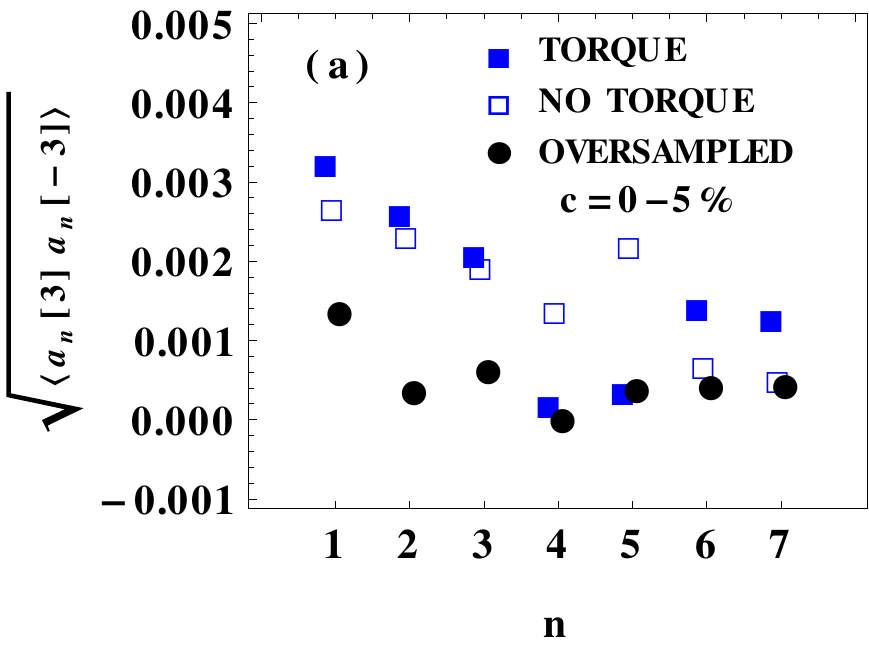}\\ 
\vspace{-8mm}
\includegraphics[width=0.45 \textwidth]{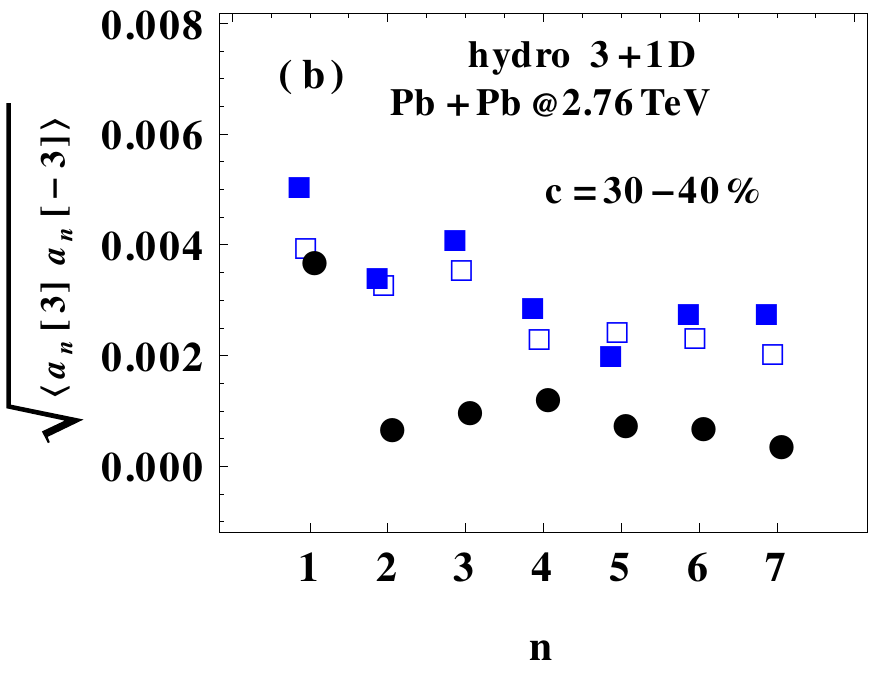}
\caption{(color online) Same as in Fig. \ref{fig:ann} but for the coefficients of the third harmonic $\sqrt{\langle a_n[3] a_n[-3]\rangle}$.
\label{fig:annv3}}
\end{figure}

Collective flow  in ultrarelativistic heavy-ion collisions causes all particles to be emitted in a correlated way, which leads to azimuthal asymmetry
in hadron distributions. 
The correlation function between two pseudorapidity  bins, constructed for multiplicity correlations as in the preceding sections, can be straightforwardly
generalized for each harmonic flow component for any two pseudorapidity bins. Various techniques are applicable here. The rapidity dependence could be decomposed into 
principal components~\cite{Bhalerao:2014mua}, but in
practice the principal component analysis may be difficult and restricted to the lowest eigenmodes. 
Alternatively, the harmonic flow correlations in pseudorapidity
can be expanded in a basis of suitable orthogonal polynomials, in full analogy to the multiplicity case. This provides insight into a different characteristic of the flow,
with possibly different sensitivity to non-flow effects than in the multiplicity correlations discussed in Sec.~\ref{sec:corr}.

Let us define the correlation coefficients for the $n$-th order harmonic flow as
\begin{equation}
\langle a_j[n] a_k[-n] \rangle = \left\langle \sum_{a\neq b}  \frac{T_i\left(\frac{\eta_a}{Y}\right)e^{i n \phi_a}}{Y \langle N(\eta_a) \rangle}
 \frac{T_k\left(\frac{\eta_b}{Y}\right)e^{-i n \phi_b}}{Y \langle N(\eta_b) \rangle}  \right\rangle  \ .
\label{eq:anmvn}
\end{equation}
In the above equation, use  we the normalization of the correlations function by ${1}/{\langle N(\eta_1) \rangle \langle N(\eta_2) \rangle}$ as in Eq.~(\ref{eq:C2}),
but the formula can be written analogously for the correlation function of flow vectors in two rapidity intervals as used in~\cite{Bhalerao:2014mua}.
Note that the linear part of the pseudorapidity dependence of the torque effect for the orientation of the flow angle~\cite{Bozek:2010vz} contributes to the  $\langle a_1[n] a_1[-n] \rangle$ coefficient.

In Figs. \ref{fig:annv2} and \ref{fig:annv3} we show the decomposition coefficients (\ref{eq:anmvn}) of the 
elliptic and triangular flow correlation at different pseudorapidities. We compare calculations using the torque and no-torque scenarios for the initial  conditions,
as in Sect. \ref{sec:hydro}. We notice that the two calculation give similar results, although in the no-torque case the odd coefficients should vanish within the statistical uncertainties. 
Our results mean that 
in the decomposition of  the flow correlations in pseudorapidity, the dominant contribution comes from  resonance decays.
The same effect has  been noticed in the analysis of factorization breaking for flow at different pseudorapidities~\cite{Bozek:2015bha} (the torque effect).

\section{Higher-order cumulants \label{sec:cumul}}

\begin{figure}[tb]
\includegraphics[width=0.45 \textwidth]{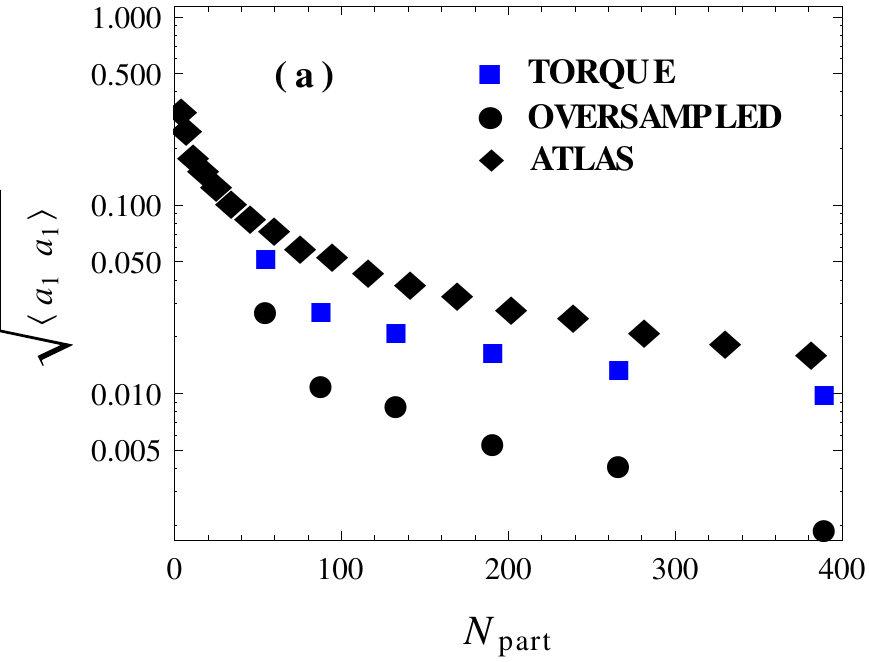}\\
\vspace{-7mm}
\includegraphics[width=0.45 \textwidth]{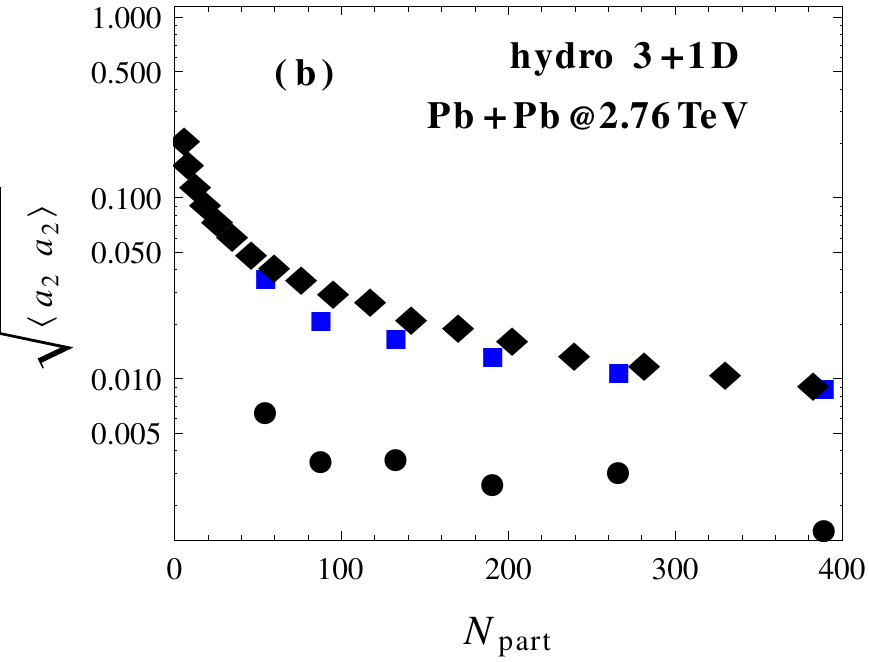} 
\caption{(color online) Coefficients $\sqrt{\langle a_1 a_1 \rangle }$ (panel a) and 
$\sqrt{\langle a_2 a_2 \rangle }$ (panel b) for the torque model and for the oversampled events for the torque case, 
plotted as functions of the number of participants (determining centrality). The ATLAS data come from Ref.~\cite{ATLAS:2015kla}.
\label{fig:anncomb}}
\end{figure}

Non-flow correlations have a significant contribution to the measured $\langle a_n a_m  \rangle$ coefficients.
In this Section we show results of an idealized calculation with the non-flow effects removed. The coefficients are calculated using {\em oversampled} 
events, where for each hydrodynamic evolution several hundreds of THERMINATOR events are generated and  combined together. That way non-flow
effects are damped. The procedure is equivalent to a Monte Carlo integration of one-particle densities in each event.
The results for $\langle a_1 a_1  \rangle$  and $\langle a_2 a_2  \rangle$  are show in Fig.~\ref{fig:anncomb}. The genuine effect due
to event by event fluctuations of the rapidity distributions is small. The coefficients from the shape fluctuations are much smaller than
correlations from resonance decays. The dependence on the rank $n$ of
 the coefficients $\langle a_n a_n \rangle$ from oversampled events 
is presented in Figs. \ref{fig:ann} to \ref{fig:annv3}. As expected in 
the torque
 scenario, involving forward-backward asymmetry, 
$\langle a_1 a_1 \rangle$ and $\langle a_1[n] a_1[-n] \rangle$ 
have the largest magnitude.

Two particle correlations from resonances are removed in higher order cumulants~\cite{Bzdak:2015dja}, while genuine correlations  
due to initial-state fluctuations of  rapidity distributions (\ref{eq:aexp}) do contribute. 
There are many possible combinations of the fourth-order cumulants that can be used for the purpose. In particular, one can define the simplest fourth-order cumulant for the 
multiplicity fluctuations as~\cite{Bzdak:2015dja}
\begin{widetext}
\begin{eqnarray}
\langle a_k^4\rangle_c &=& \left\langle  {\sum_{a,b,c,d}}^{'}  \frac{T_k\left(\frac{\eta_a}{Y}\right)}{Y \langle N(\eta_a) \rangle}
 \frac{T_k\left(\frac{\eta_b}{Y}\right)}{Y \langle N(\eta_b) \rangle} \frac{T_k\left(\frac{\eta_c}{Y}\right)}{Y \langle N(\eta_c) \rangle} \frac{T_k\left(\frac{\eta_d}{Y}\right)}{Y \langle N(\eta_d) \rangle}  \right\rangle 
- 3 \left\langle  {\sum_{a,b}}^{'}  \frac{T_k\left(\frac{\eta_a}{Y}\right)}{Y \langle N(\eta_a) \rangle}
 \frac{T_k\left(\frac{\eta_b}{Y}\right)}{Y \langle N(\eta_b) \rangle}  \right\rangle \ ,
\label{eq:cumu4}
\end{eqnarray}
where the subscript $c$ stands for the connected part and the prime denotes summation over different particles.  
For the flow correlations in pseudorapidity, the most general cumulant is of the form
\begin{eqnarray}
\langle a_{i_1}[m_1] \dots  a_{i_n}[m_n] \rangle_c = \left\langle \sum_{a_1,\dots, a_n}  \frac{T_{i_1}\left(\frac{\eta_{a_1}}{Y}\right)e^{i m_1 \phi_{a_1}}}{Y \langle N(\eta_{a_1}) \rangle} \dots
 \frac{T_{i_n}\left(\frac{\eta_{a_n}}{Y}\right)e^{i m_n \phi_{a_n}}}{Y \langle N(\eta_{a_n})\rangle }  \right\rangle_c   \ , 
\label{eq:anmvcum}
\end{eqnarray}
with $\sum_{k=1}^n m_k=0$. 
The simplest fourth-order cumulants are
\begin{eqnarray}\label{eq:vcumu4}
&& \langle a_k[n]a_k[n]a_k[-n]a_k[-n]\rangle_c=  \\ &&~~
\left\langle  {\sum_{a,b,c,d}}^{'}  \frac{T_k\left(\frac{\eta_a}{Y}\right)e^{i n \phi_{a}}}{Y \langle N(\eta_a) \rangle}
 \frac{T_k\left(\frac{\eta_b}{Y}\right)e^{i n \phi_{b}}}{Y \langle N(\eta_b) \rangle} \frac{T_k\left(\frac{\eta_c}{Y}\right)e^{-i n \phi_{c}}}{Y \langle N(\eta_c) \rangle}
 \frac{T_k\left(\frac{\eta_d}{Y}\right)e^{-i n \phi_{d}}}{Y \langle N(\eta_d) \rangle}  \right\rangle - 2 \left\langle  {\sum_{a,b}}^{'}  \frac{T_k\left(\frac{\eta_a}{Y}\right)e^{i n \phi_{a}}}{Y \langle N(\eta_a) \rangle}
 \frac{T_k\left(\frac{\eta_b}{Y}\right)e^{-i n \phi_{b}}}{Y \langle N(\eta_b) \rangle}  \right\rangle \ . \nonumber
\end{eqnarray}
\end{widetext}

We have attempted to compute the fourth-order cumulants (\ref{eq:cumu4}) and (\ref{eq:vcumu4}) in our simulation, however, with the available statistics 
(20000 events in each centrality class) the statistical errors are too large, of the  same order as the square of the second order cumulant.
The application of the cumulant method \cite{Bzdak:2015dja} is possible on large-statistics experimental data and the results could be compared
to model calculations using one-particle densities (such as the results for the oversampled events presented above) that neglect the non-flow effects.

\section{Conclusions}

We have checked the predictions of a realistic simulation based on viscous 3+1-dimensional hydrodynamics for the 
two-particle correlations in pseudorapidity, as measured by the ATLAS collaboration~\cite{ATLAS:2015kla} and found that the 
correlation from the resonance decays, formed at a late stage of the evolution, produce significant effects. In particular, their contribution
to the coefficients $\langle a_n a_m \rangle$  in the expansion of the correlation function in the Legendre basis take 60-70\% of the experimental values. 

While our model incorporates only some possible sources of correlations (those from the torque effect and the resonance decays), it shows their 
relevance in the analyses. Other, not incorporated  non-flow effects, include the hadron production from jets, local current conservation, or 
additional sources of rapidity fluctuations in the initial state~\cite{Bozek:2015bna}.

On the methodological level, we have proposed a new way to compute the  $\langle a_n a_m \rangle$ coefficients, independent of the binning in pseudorapidity, and applied it in our 
simulations. Also, we have developed a double expansion of the correlation function in the azimuth and pseudorapidity, which allows to probe 
and quantify the rapidity correlations between harmonics of the collective flow. We have found that in our model these quantities are also dominated by 
non-flow effects.
Our method can be used for higher-order averages of the orthogonal polynomials, in particular for cumulants.
This offers a way of eliminating the non-flow effects \cite{Bzdak:2015dja}, but requires very large statistics, which, fortunately, is available in the experiments.
These measures could be compared to model calculations with oversampled events, where sufficient statistics can be achieved.

We note that a study using similar methods and leading to similar
results has been independently and simultaneously presented in
Ref.~\cite{Monnai:2015sca}.

\begin{acknowledgments}

We thank Jiangyong Jia  for clarification of the experimental analysis. Research supported by the Polish Ministry of Science and Higher Education (MNiSW), by the National
Science Center grants DEC-2012/05/B/ST2/02528 and DEC-2012/06/A/ST2/00390, as well as by PL-Grid Infrastructure.

\end{acknowledgments}

\bibliography{../../hydr}

\end{document}